\documentclass[aps,pre,twocolumn,superscriptaddress,amsmath,amssymb]{revtex4-2} 

\usepackage{amssymb,amsmath,amsbsy}
\usepackage{upgreek}
\usepackage{cancel}
\usepackage{mathdots}
\usepackage{stackrel}
\usepackage{enumerate}
\usepackage{mathrsfs}
\usepackage{graphicx}
\usepackage{float}
\usepackage{mathrsfs}
\usepackage{physics}
\usepackage{listings}
\usepackage{hyperref}
\usepackage{pgf}
\usepackage{tikz}
\usetikzlibrary{arrows.meta, positioning, calc, 3d, decorations.pathmorphing}
\usetikzlibrary{decorations.pathmorphing, arrows.meta, positioning}
\usetikzlibrary{external}
\usepackage{polynom}
\usepackage{soul}

\begin{document}

\newcommand\chau{\affiliation{ 
 Charles University,  
 Faculty of Mathematics and Physics, 
 Department of Macromolecular Physics, 
 V Hole{\v s}ovi{\v c}k{\' a}ch 2, 
 CZ-180~00~Praha, Czech Republic 
}}

\title{Strong Violation of the Thermodynamic Uncertainty Relation in a Minimal Autonomous Heat Engine}

\author{Enrique P. Cital}\chau
\author{Viktor Holubec}\email{viktor.holubec@mff.cuni.cz}\chau
\date{\today}
\begin{abstract}
Thermodynamic uncertainty relations (TURs) impose a universal trade-off between current precision and entropy production in autonomous steady states, constraining in particular the power, efficiency, and constancy of heat engines. We demonstrate strong violations of the long-time TUR in a minimal autonomous heat engine composed of a discrete ratchet generating work against a constant bias and an underdamped harmonic oscillator acting as an internal stochastic control. In the regime of time-scale separation, the model becomes exactly solvable and yields a closed analytical expression for the TUR ratio, where the influence of the continuous degree of freedom is fully captured by the Fano factor of oscillator zero crossings. We show that increasingly deterministic internal control drives the TUR ratio arbitrarily close to zero while the engine operates near maximal current and efficiency. In an appropriate limit, the model reduces to the classical pendulum-clock system of Pietzonka, Phys. Rev. Lett. \textbf{128}, 130606 (2022).
\end{abstract}

\maketitle

\section{Introduction}
Heat engines (HE) are the canonical example of how temperature gradients can be converted into directed motion. In the macroscopic world, their output power is effectively deterministic. At microscopic scales, however, thermal noise strongly affects engine performance, rendering the output power subject to significant fluctuations~\cite{Holubec_2022}. This aspect is particularly relevant today, as current experimental capabilities allow the realization and precise control of such microscopic HEs~\cite{SingelAtomHE2016,ColloidalReview,adiabatic_realization,Ciliberto2017}.

Due to the presence of fluctuations, the classical power--efficiency trade-off that governs macroscopic HEs extends at microscopic scales to a three-way trade-off between power, efficiency, and power fluctuations. A central theoretical result related to this trade-off is provided by thermodynamic uncertainty relations (TURs)~\cite{Seifert2012RPP,BaratoSeifert2015PRL,GingrichEtAl2016PRL,HorowitzGingrich2020NatPhys}. In its original form, TUR is valid for dynamics with a simple time-reversal structure—such as time-homogeneous Markov jump processes and overdamped Langevin motion. It reads
\begin{equation}
\mathcal{T} = \frac{\mathrm{Var}\!\left[J(t)\right]}{\left\langle J(t)\right\rangle^{2}} \frac{\sigma t}{2}
\ge
1,
\label{eq:TUR_intro}
\end{equation}
where $J(t)$ denotes a stationary stochastic current averaged over a measurement time $t$, $\sigma$ is the stationary entropy production rate, all averages are taken over realizations of the stochastic noise, and we set the Boltzmann constant $k_{\rm B} = 1$.

Since the TUR~\eqref{eq:TUR_intro} holds for arbitrary autonomous systems with a given time-reversal symmetry, it is also valid for corresponding autonomous HEs. In this case, it can be rewritten in the form~\cite{pietzonka2018universal}
\begin{equation}
    \mathcal{T} = 
    \frac{\mathrm{Var}[P(t)]\, t}{\langle P(t) \rangle}
    \frac{\eta - \eta_{\rm C}}{2 T_- \eta} 
    \ge 1,
    \label{eq:PEC}
\end{equation}
which quantifies the so-called power--efficiency--constancy (PEC) trade-off between the average output power $\langle P(t) \rangle$, its variance $\mathrm{Var}[P(t)]$ (or constancy when multiplied by time), and the efficiency $\eta$.

The time-reversal symmetry necessary for the validity of TUR~\eqref{eq:TUR_intro} can be broken in autonomous systems when variables that are odd under time reversal, such as magnetic fields~\cite{Chun2019}, velocities~\cite{Pietzonka2022}, or quantum degrees of freedom~\cite{PhysRevLett.126.010602}, play a role. Similarly, the symmetry is broken in systems driven by external time-dependent control~\cite{HolubecMaxPower2018}. While extended variants of TUR~\eqref{eq:TUR_intro} can be derived for such systems, these bounds involve additional kinetic or dynamical contributions beyond entropy production~\cite{LeeParkPark2019PRE,FuGingrich2022PRE}. As a result, they do not admit a simple rewriting in terms of the PEC trade-off~\eqref{eq:PEC}.

Although the TUR—and consequently the PEC trade-off—is known to break down in autonomous underdamped and quantum systems, explicit examples are readily available only for short measurement times $t$~\cite{Fischer2020,CitalHolubec2026InertiaTames}. 
In the quantum domain, an autonomous HE whose performance violates the PEC trade-off at long times $t$ has been reported only recently~\cite{Meier2025}. 
It consists of $n$ coupled quantum systems arranged in a ring. Upon optimizing the coupling constants between the subsystems, the engine exhibits a TUR ratio $\mathcal{T}$ that, for large $n$, scales as $\log n / n^{4/3}$. 
The experimental realization of such a system, however, remains highly nontrivial.

An autonomous underdamped HE that violates the long-time TUR was introduced around the same time~\cite{CitalHolubec2026InertiaTames}. 
Building on an ingenious setup involving a dragged underdamped particle whose current was shown to violate the TUR~\cite{Pietzonka2022}, we constructed an autonomous HE comprising two continuous underdamped degrees of freedom and one discrete degree of freedom. 
The discrete variable represents the engine component, with its temperature controlled by one of the underdamped modes acting as a stochastic internal drive. 
The output current is generated in the second underdamped degree of freedom. 
Remarkably, when the system operates under analytically predictable resonant conditions, the TUR ratio $\mathcal{T}$ attains values on the order of $10^{-3}$, comparable to the numerical uncertainty of the Brownian dynamics simulations. 
Although the proposed setup is likely more experimentally feasible than its quantum counterpart, the nonlinear coupling between the underdamped degrees of freedom has so far precluded a complete analytical characterization of the mechanism responsible for this pronounced violation.

In the present work, we introduce a minimal autonomous HE consisting of a single continuous underdamped degree of freedom coupled to a single discrete degree of freedom that violates the long-time TUR. In contrast to the previous setup, the output current is generated in the discrete variable, while the underdamped mode serves as a noisy, time-dependent internal control. 
The construction is again inspired by the model of Ref.~\cite{Pietzonka2022}. In the limit where the dynamics of the continuous degree of freedom is much slower than that of the discrete one, the model becomes exactly solvable due to time-scale separation. 
This solvability enables us to analytically determine the conditions under which the TUR ratio $\mathcal{T}$ is minimized, thereby identifying regimes of highly precise and reliable autonomous HEs or stochastic clocks~\cite{BaratoSeifert2016PRX,ErkerEtAl2017PRX}.

\begin{figure*}
\centering
\includegraphics[width=\textwidth]{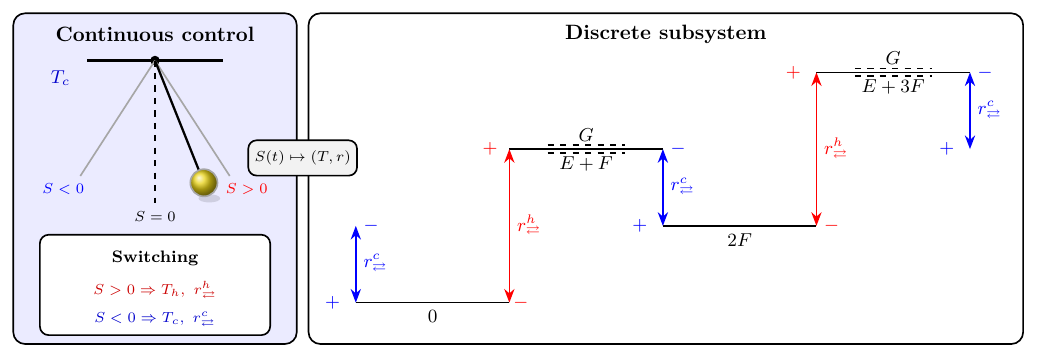}
\caption{Sketch of the model. The underdamped harmonic oscillator (left) controls the temperature of the bath and the transition rates in the discrete subsystem (right). Specifically, when the oscillator coordinate is positive (negative), the discrete subsystem is in contact with the hot (cold) bath at temperature $T_h$ ($T_c$), and the red (blue) transitions are allowed. The signs $\mp$ indicate the direction of the positive current. The dashed energy levels are $G$-fold degenerate, and the labels below the levels denote their energies.}
\label{fig:Fig2}
\end{figure*}

The remainder of the paper is organized as follows. 
In Sec.~\ref{sec:fullmodel}, we introduce the model and its thermodynamic description. 
In Sec.~\ref{sec:solution}, we derive analytical expressions for the current, its fluctuations, and the TUR ratio, which are used in Sec.~\ref{sec:violation} to identify the parameter regime of maximal TUR violation. 
We conclude in Sec.~\ref{sec:Conclusion}.

\section{Model}
\label{sec:fullmodel}

We consider a system composed of one continuous ``control'' degree of freedom and one discrete ``HE'' degree of freedom, as sketched in Fig.~\ref{fig:Fig2}. 
The control modulates the temperature of the bath and the transition rates in the discrete subsystem, thereby inducing a current against an external force. 
In the present section, we describe the stochastic dynamics and thermodynamics of the system.

\subsection{Continuous subsystem}

We model the control subsystem as an equilibrium underdamped harmonic oscillator. In dimensionless form, it is described by the Langevin equation
\begin{equation}
\ddot{S}(t) = -S(t) - \gamma\,\dot{S}(t) + \sqrt{2\gamma}\,\xi(t),
\label{eq:clock_u}
\end{equation}
where $\xi(t)$ is a normalized, unbiased Gaussian white noise. The dynamics is governed by the dimensionless parameter $\gamma = \Gamma/(m\omega)$, defined in terms of the friction coefficient $\Gamma$, the oscillator mass $m$, and its natural angular frequency $\omega$. Time is measured in units of $\omega^{-1}$ and length in units of $\sqrt{T_c/(m\omega^2)}$, where $T_c$ denotes the reservoir temperature.

As discussed below, we assume that the coupling between the oscillator and the discrete system is sufficiently weak that the backaction of the discrete system on the continuous one can be neglected. Under such conditions, the oscillator relaxes to an equilibrium steady state characterized by vanishing entropy production.


\subsection{Discrete subsystem}

We assume that the discrete degree of freedom performs a random walk in the one-dimensional energy landscape depicted in Fig.~\ref{fig:Fig2}. The even states, with energies proportional to $iF$, where $i=\dots,-2,0,2,\dots$ denotes the discrete position on the lattice, are nondegenerate. 
The odd states, with energies $E+iF$ for $i=\dots,-1,1,\dots$, have degeneracy $G$. The entire energy landscape is tilted by the bias $F$, and the HE performs work when the particle jumps uphill against this bias.

The transition rates between the individual energy levels and the temperature $T$ of the reservoir in contact with the discrete degree of freedom are controlled by the continuous variable $S(t)$. When $S(t) > 0$, $T = T_h$ and only the red transitions labeled with superscript $h$ in Fig.~\ref{fig:Fig2} are allowed. When $S(t) < 0$, $T = T_c < T_h$, and only the blue transitions with superscript $c$ can occur. Hence, at any given time, the discrete degree of freedom behaves as a two-level system.

Per half-oscillation of the oscillator, the coordinate $i$ in the discrete landscape can change by at most one. This mechanism suppresses possible current fluctuations and cannot be implemented entirely for free~\cite{Pietzonka2022,CitalHolubec2026InertiaTames}. Nevertheless, the associated cost can be made negligible, for example, by introducing very thin energy barriers between the states~\cite{Pietzonka2022} or by tuning the transitions to specific wavelengths in a quantum-optical setup, where the oscillator position controls which heat source, characterized by a tailored radiation spectrum, interacts with the system at a given time. Here, we neglect this cost entirely for analytical simplicity. For more complex setups in which the coupling is treated explicitly, see Refs.~\cite{Pietzonka2022,CitalHolubec2026InertiaTames}.

We assume that the transition rates obey the detailed-balance condition with respect to the temperature of the reservoir currently coupled to the discrete subsystem and are given by
\begin{eqnarray}
r^{h}_{\rightleftarrows} &=& k^{h}(S)\,{\rm e}^{\mp \frac{1}{T_{h}}\left(\frac{E+F}{2}\right)}, \label{eq:h_trans}\\
r^{c}_{\rightleftarrows} &=& k^{c}(S)\,{\rm e}^{\pm \frac{1}{T_{c}}\left(\frac{E-F}{2}\right)}, \label{eq:c_trans}
\end{eqnarray}
where the arrows in the lower index denote the transition directions shown in Fig.~\ref{fig:Fig2}. 
The upper indices indicate which heat reservoir is coupled to the discrete subsystem when the corresponding transition rates are nonzero, i.e.,
\begin{equation}
\begin{aligned}
k^{h}(S>0)&=k_0, &\qquad k^{c}(S>0)&=0,\\
k^{h}(S<0)&=0,   &\qquad k^{c}(S<0)&=k_0,
\end{aligned}
\end{equation}
where $k_0$ is a constant attempt rate. 

Due to the degeneracy of the odd states, the total transition rates from lower to higher energy are $G r^{h}_{\rightarrow}$ at $T=T_h$ and $G r^{c}_{\leftarrow}$ at $T=T_c$. The transition rates to lower-energy states are those given in Eqs.~\eqref{eq:h_trans} and~\eqref{eq:c_trans}.


\subsection{Thermodynamics}

As mentioned above, in the steady state the continuous subsystem undergoes equilibrium oscillations and cyclically controls the discrete subsystem. The latter can therefore be viewed as a cyclically driven ratchet with stochastic durations of the individual cycle branches.

Denoting by $I(t)\in\{\dots,-1,0,1,\dots\}$ the stochastic state of the discrete subsystem at time $t$ in the steady state, the time-averaged stochastic current over the interval $[0,t]$ is defined as
\begin{equation}
J(t) = \frac{I(t) - I(0)}{t}.
\end{equation}
To characterize the average thermodynamic properties of the HE, we consider the mean value of $J(t)$ integrated over a single half-oscillation of the control subsystem, denoted by $j_0$. 
Since between successive sign changes of $S(t)$ the discrete subsystem behaves as a two-level system, it follows that $j_0 \in [-1,1]$.

Upon crossing one unit cell of the energy landscape in Fig.~\ref{fig:Fig2}, the discrete subsystem absorbs an amount of energy $E+F$ from the hot bath and releases an amount of energy $E-F$ to the cold bath. Because the subsystem is in contact with each heat reservoir for half of the cycle duration, the average heat exchanged with the reservoirs per half-oscillation is given by
\begin{eqnarray}
    q_h &=& \frac{1}{2}(E+F)\, j_0,\\
    q_c &=& -\frac{1}{2}(E-F)\, j_0.
\end{eqnarray}
The average output work per half-oscillation and the efficiency of the discrete engine are then
\begin{eqnarray}
    w &=& q_h + q_c = F j_0,\\
    \eta &=& \frac{w}{q_h} = \frac{2F}{E+F}.
\end{eqnarray}
Notably, the efficiency is independent of the dynamical details encapsulated in $j_0$, which cancels between the numerator and denominator. Assuming a positive slope of the energy landscape, $F > 0$, Carnot's theorem, $\eta \le \eta_{\rm C} = 1 - T_c/T_h$, implies the condition
\begin{equation}
  \frac{E - F}{E + F} \ge \frac{T_c}{T_h},
  \label{eq:2ndlaw}
\end{equation}
which must be satisfied for the engine to deliver positive output work and thus sustain a positive average current.

Since the stochastic output power $P(t) = F J(t)$ is directly proportional to the current $J(t)$, the TUR~\eqref{eq:TUR_intro} can in the present case be rewritten in the form of the power–efficiency–constancy trade-off~\eqref{eq:PEC}. To see this, note that the average entropy produced by the whole system per half-oscillation in the steady state is given by
(the continuous degree of freedom operates without dissipation)
\begin{equation}
    \sigma_0  = -\frac{q_h}{T_h} - \frac{q_c}{T_c} = \Delta j_0,
    \label{eq:stoj}
\end{equation}
where the \emph{cycle affinity} reads
\begin{equation}
\Delta \equiv \frac{1}{2}\left(\frac{E-F}{T_c}-\frac{E+F}{T_h}\right)
=
\frac{F}{T_c}\,
\frac{\eta_{\rm C} - \eta}{\eta}.
\label{eq:delta}
\end{equation}
Hence, by writing the entropy production rate $\sigma$ in Eq.~\eqref{eq:TUR_intro} as $\sigma = \Delta \langle J(t) \rangle$ and using $P(t) = F J(t)$, we readily obtain Eq.~\eqref{eq:PEC}. Note that the second-law inequality implies $\Delta > 0$ and thus $\sigma > 0$, as required.

\section{Approximate analytical solution}
\label{sec:solution}

Per half-oscillation, the displacement $j_0$ increases by one if the subsystem transitions from a $-$ to a $+$ state in the energy landscape shown in Fig.~\ref{fig:Fig2}, and decreases by one for the reverse transition.

For the remainder of this section, we assume time-scale separation, such that the discrete subsystem relaxes sufficiently fast for the populations of the two-level system to equilibrate at the instantaneous reservoir temperature between successive sign changes of $S(t)$. In practice, this requires the intrinsic attempt rate $k_0$ to be much larger than the oscillator frequency $\omega$.

\begin{figure}
    \centering
  \includegraphics[width=1.0\columnwidth]{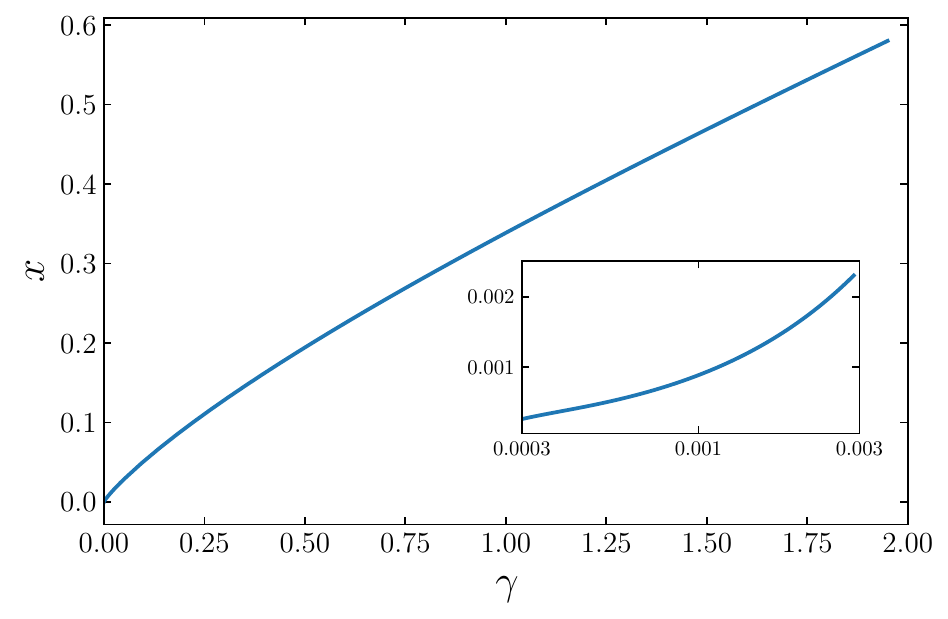}
  \caption{
Fano factor $x$ of the harmonic oscillator, defined in Eq.~\eqref{eq:x_def_fullmodel}, as a function of the reduced friction coefficient $\gamma$.}
   \label{fig:xVSgammadimensionless}
\end{figure}

The equilibrium occupation probability of the $+$ states is given by
\begin{equation}
p^{h}_{+} = \frac{G r^{h}_{\rightarrow}}{G r^{h}_{\rightarrow}+r^{h}_{\leftarrow}},
\end{equation}
for $S>0$, and
\begin{equation}
p^{c}_{+} = \frac{r^{c}_{\rightarrow}}{r^{c}_{\rightarrow}+G r^{c}_{\leftarrow}},
\end{equation}
for $S<0$.
The corresponding occupation probabilities of the $-$ states are $p^{h}_{-}=1-p^{h}_{+}$ and $p^{c}_{-}=1-p^{c}_{+}$, respectively.

The hot and cold transitions induce two independent biased random walks with probabilities $p^{h,c}_{+}$ for right jumps and $p^{h,c}_{-}$ for left jumps, each acting over one half of the oscillation period. The corresponding average displacements read
\begin{equation}
j^{h,c}_{0} \equiv \frac{1}{2}\left(p^{h,c}_{+}-p^{h,c}_{-}\right),
\label{eq:curr_hc}
\end{equation}
and the associated variances are given by
\begin{equation}
v^{h,c}_0
= \frac{1}{2}\left[p^{h,c}_{+}+p^{h,c}_{-}-\left(p^{h,c}_{+}-p^{h,c}_{-}\right)^{2}\right].
\end{equation}

Substituting the transition rates~\eqref{eq:h_trans} and~\eqref{eq:c_trans} into Eq.~\eqref{eq:curr_hc}, the total average displacement per half-oscillation, $j_0 = j^h_{0}+j^c_{0}$, is found to be
\begin{equation}
j_0 = -1 + \frac{G}{{\rm e}^{(F+E)/T_h} + G} + \frac{1}{1 + {\rm e}^{(F-E)/T_c} G}.
\label{eq:curr_swing}
\end{equation}
The maximal negative displacement, $j_0 = -1$, is naturally achieved in the limit of infinite bias, $F \to \infty$. Interestingly, a nonzero degeneracy of the excited states also enables the system to reach the maximal positive displacement, $j_0 = 1$, in the limit of an infinite temperature difference and infinite degeneracy, provided that ${\rm e}^{(F-E)/T_c}G \to 0$.

Since the two random walks are independent, the overall displacement variance is
\begin{multline}
v_0 = v^{h}_0 + v^{c}_0
=
2\left[p^{h}_{+}+p^{c}_{+}-\left(p^{h}_{+}\right)^{2}-\left(p^{c}_{+}\right)^{2}\right] \\
= 2 j_0 \left(  \coth \Delta-j_0 \right).
\label{eq:v0}
\end{multline}

To convert the displacement into a current, it must be multiplied by the average number of half-oscillations per unit time, $\langle \dot{N} \rangle$. A half-oscillation corresponds to a sign change of $S(t)$. 
For long measurement times involving many oscillations, the stochastic number $N(t)$ of sign changes of $S(t)$ within a time window of duration $t$ in the steady state can be approximated as a Gaussian random variable with mean and variance given by
\begin{equation}
\langle N(t)\rangle \simeq \langle \dot{N}\rangle\,t,
\qquad
{\rm Var}\!\left[N(t)\right]\simeq 2D_N\,t.
\label{eq:CLT_ticks_full}
\end{equation}
Both the mean rate of sign changes, $\langle \dot{N} \rangle$, and the corresponding diffusion coefficient, $D_N$, can be calculated analytically~\cite{Pietzonka2022}.

\begin{figure}
    \centering
  \includegraphics[width=1.0\columnwidth]{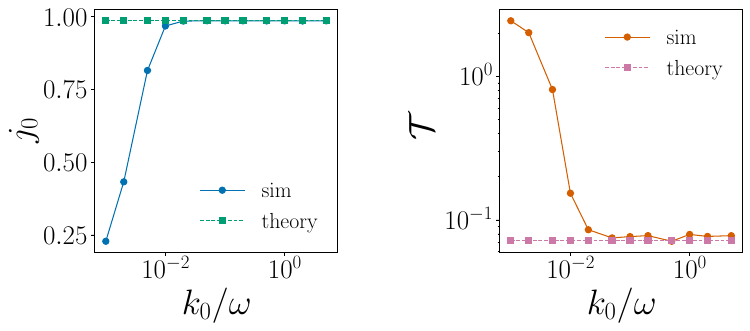}
   \caption{Average displacement per half-oscillation, $j_0$ (left), and TUR ratio, $\mathcal{T}$ (right), as functions of $k_0/\omega$ for $E=3$, $F=2$, $G=150$, $T_h=1000$, $T_c=0.1$, and $\gamma=10^{-3}$ ($x = 0.009$) in reduced units. Circles (solid lines) are obtained from Brownian dynamics simulations of the full model, and squares (dashed lines) from the approximate analytical formulas in Eqs.~\eqref{eq:curr_swing} and~\eqref{eq:T_general_label}.}
   \label{fig:NUMvsANAL}
\end{figure}

The average integrated current in the discrete subsystem over a long time $t$ in the steady state is given by
\begin{equation}
\langle J(t) \rangle = j_{0} \langle N(t)\rangle  = j_{0} \langle \dot{N} \rangle \, t.
\end{equation}
Since the oscillations are independent of the discrete dynamics, the variance of the integrated current reads~\cite{Pietzonka2022}
\begin{equation}
\mathrm{Var}\!\left[ J(t) \right] =  v_0 \langle \dot{N}\rangle\, t + j_0^2\, \mathrm{Var}\!\left[ N(t)\right].
\label{eq:varfull}
\end{equation}

Substituting the average current and its variance from Eqs.~\eqref{eq:curr_swing}--\eqref{eq:varfull}, together with the entropy production rate $\sigma = \sigma_0 \langle \dot{N} \rangle$, where $\sigma_0$ is given by Eq.~\eqref{eq:stoj}, into the TUR ratio in Eq.~\eqref{eq:TUR_intro}, we obtain
\begin{equation}
\mathcal{T}
= \Delta \left( \coth \Delta -(1-x)\,j_0 \right),
\label{eq:T_general_label}
\end{equation}
where the entire effect of the oscillator degree of freedom is captured by the Fano factor
\begin{equation}
x \equiv \frac{D_N}{\langle \dot N\rangle}.
\label{eq:x_def_fullmodel}
\end{equation}
For the harmonic oscillator~\eqref{eq:clock_u}, the Fano factor can be evaluated analytically~\cite{Pietzonka2022}. The result is shown in Fig.~\ref{fig:xVSgammadimensionless}.

To test the analytical predictions derived in this section, we solved the system dynamics numerically using Brownian dynamics simulations for various values of the attempt rate $k_0$ and the oscillator frequency $\omega$. A representative result is shown in Fig.~\ref{fig:NUMvsANAL}. It demonstrates that the average displacement per half-oscillation, $j_0$, and the TUR ratio, $\mathcal{T}$, obtained numerically converge to the analytical predictions in Eqs.~\eqref{eq:curr_swing} and~\eqref{eq:T_general_label} as the ratio $k_0/\omega$ increases.

\section{TUR violation}
\label{sec:violation}

The expression~\eqref{eq:T_general_label} for the TUR ratio $\mathcal{T}$ is symmetric with respect to the direction of the current and therefore remains unchanged when the discrete system transitions along the bias $F$. To see this, it suffices to note that the second law implies $\sigma_0 = \Delta j_0 > 0$ under all circumstances; hence, $\Delta$ and $j_0$ necessarily have the same sign. To investigate the behavior of $\mathcal{T}$ over the full parameter space, it is therefore sufficient to focus on the heat-engine regime of operation ($j_0>0$).

The TUR ratio is proportional to the cycle affinity $\Delta$. Hence, it may appear to vanish in the limit $\Delta \to 0$, when the HE efficiency approaches the Carnot efficiency, $\eta \to \eta_{\rm C}$, and the system operates reversibly. Expanding $\mathcal{T}$ to first order in $\Delta$, we obtain
\begin{equation}
\mathcal{T}
=
1
-
(1-x)\,j_0\,\Delta.
\label{eq:TUR_eta_expand_general}
\end{equation}
In the strict reversible limit $\Delta \to 0$, the TUR therefore saturates at $\mathcal{T} = 1$. Nevertheless, the expansion shows that the TUR can be weakly violated ($\mathcal{T}<1$) for small $\Delta$, provided that the oscillator Fano factor is sufficiently small, $x < 1$.

The violation of the TUR becomes stronger as the oscillator dynamics becomes more deterministic. In the limit of deterministic control, $x \to 0$, we obtain
\begin{equation}
\mathcal{T}  =\Delta\bigl[\coth \Delta - j_0\bigr].
\label{eq:ratchet_x0}
\end{equation}
In this limit, the TUR can be violated for any positive $j_0$, provided that the affinity is sufficiently small. The maximal violation is achieved at the maximal current ($j_0 = 1$), in which case the TUR is violated for all positive values of $\Delta$, and the TUR ratio $\mathcal{T}$ converges to zero as $\Delta \to \infty$.

\begin{figure}
  \centering
  \includegraphics[width=0.5\textwidth]{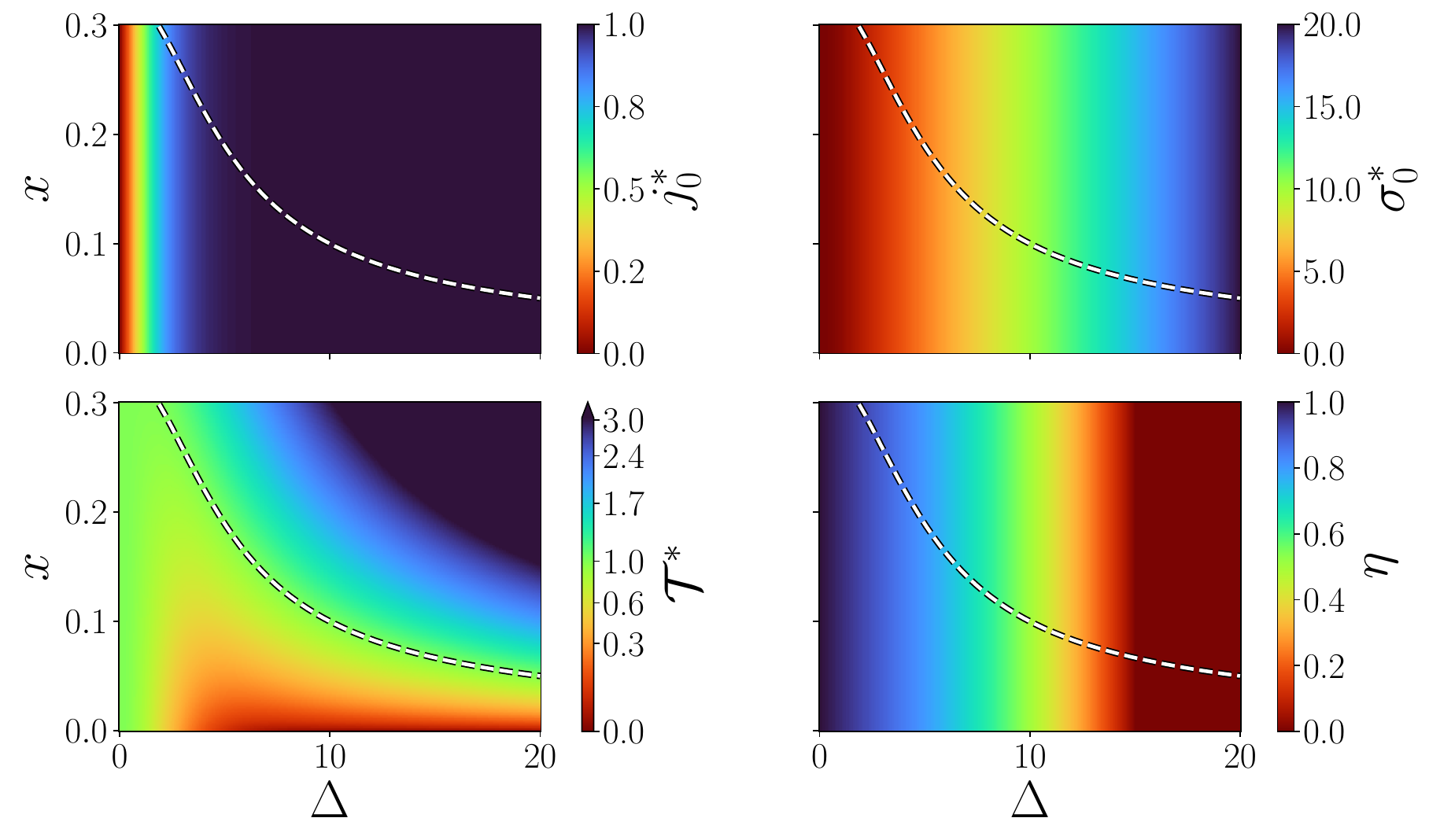}
  \caption{Average displacement per half-oscillation, $j_0^\star$, and the corresponding entropy production, $\sigma_0^\star$, together with the TUR ratio, $\mathcal{T}^\star$, and the HE efficiency, $\eta$, as functions of the cycle affinity $\Delta$ and the control precision (Fano factor) $x$ in the limit $T_h \to \infty$ and for $G = G^\star$ as defined in Eq.~\eqref{eq:Gstar_jmax}. Below the white dashed line, the TUR is violated ($\mathcal{T}^\star < 1$). Parameters: $E=3$ and $T_c=0.1$ in reduced units.
  }
\label{fig:thermodynamic_characterization}
\end{figure}

Let us now investigate the possibility of strong TUR violations for arbitrary control precision $x$. To reduce the number of model parameters, we focus on the regime of high hot-reservoir temperatures, $F+E \ll T_h$. In this limit, the probability that the discrete system transitions during the ``hot'' phase of the HE cycle in the direction of positive current is maximized, thereby suppressing fluctuations. For large degeneracy, these transitions can even become effectively deterministic. Indeed, a numerical analysis of Eq.~\eqref{eq:T_general_label} shows that in the heat-engine regime, increasing $T_h$ reduces the TUR ratio $\mathcal{T}$.

In the limit $T_h \to \infty$, the cycle affinity~\eqref{eq:delta} reduces to
\begin{equation}
\Delta = \frac{E-F}{2T_c}
\label{eq:deltaThinf}
\end{equation}
and the displacement per half-oscillation simplifies to
\begin{equation}
j_0
=\frac{(1-e^{-2\Delta})\,G}{(1+G)(1+e^{-2\Delta}G)}.
\label{eq:j0_Thinf}
\end{equation}
Since the degeneracy $G$ enters the TUR ratio only through $j_0$, and since the TUR ratio is minimized by maximizing $j_0$ at fixed $\Delta$ and $x$, we maximize $j_0$ with respect to $G$. The optimal degeneracy and the corresponding displacement are
\begin{equation}
G^\star=e^{\Delta},
\qquad
j_{0}^\star= \tanh\!\left(\frac{\Delta}{2}\right).
\label{eq:Gstar_jmax}
\end{equation}
The optimized displacement $j_0^\star$ varies between $-1$ and $1$ as the affinity $\Delta$ ranges from $-\infty$ to $\infty$, and it approaches $0$ in the reversible limit $\Delta \to 0$.

The TUR ratio~\eqref{eq:T_general_label} in the maximum-current regime reads
\begin{equation}
\mathcal{T}^\star
=\frac{\Delta}{\sinh \Delta}
+x\,\Delta\,\tanh\!\left(\frac{\Delta}{2}\right).
\label{eq:Tstar_label}
\end{equation}
This expression is identical to the TUR ratio derived for the classical pendulum-clock model in Ref.~\cite{Pietzonka2022}. Our model reduces to the pendulum-clock model in the limit $E \to 0$, $T_h \to T_c$, and $G \to 1$, where the HE does not produce work and the current is driven solely by the bias $F$. Perhaps surprisingly, the present analysis shows that the TUR ratios in the two models also coincide in the optimal HE limit $G \to G^\star$ and $T_h \to \infty$.

\begin{figure}
    \centering
  \includegraphics[width=1.0\columnwidth]{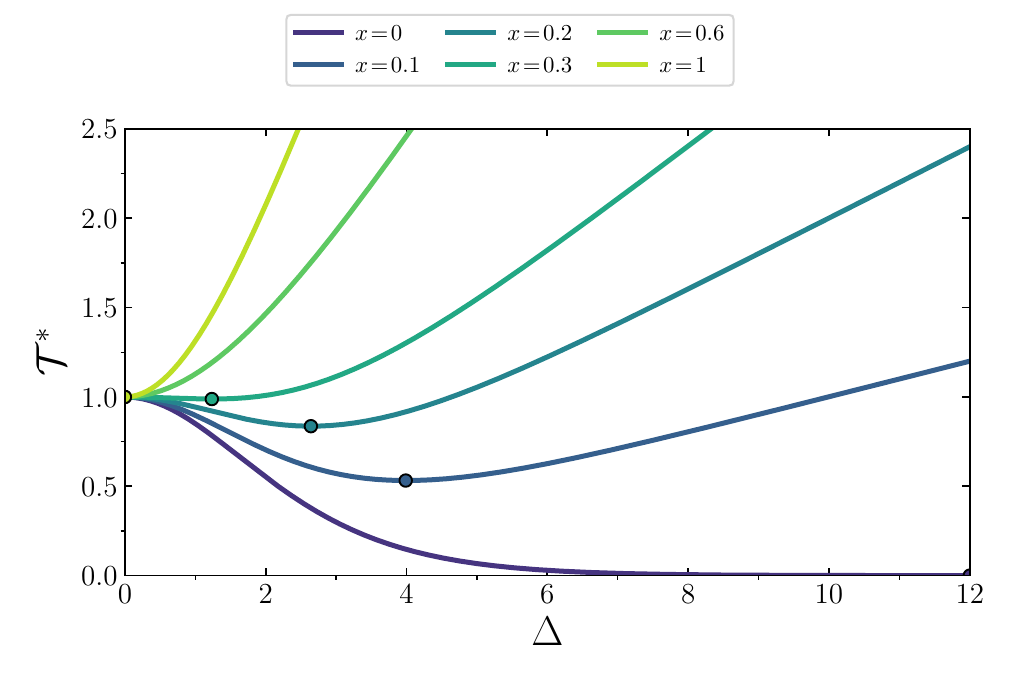}
  \caption{TUR ratio, $\mathcal{T}^{\star}$, as a function of the cycle affinity $\Delta$ in the limit $T_h \to \infty$ and for $G = G^\star$ as defined in Eq.~\eqref{eq:Gstar_jmax}, shown for six values of the control precision (Fano factor) $x$.}
   \label{fig:TURratchetminima}
\end{figure}

In Fig.~\ref{fig:thermodynamic_characterization}, we present the main characteristics of the HE as functions of the cycle affinity $\Delta$ and the control accuracy $x$: the average displacement and entropy produced per half-oscillation, $j_0$ and $\sigma_0$, which determine the average current $\langle J \rangle$ and entropy production rate $\sigma$, as well as the TUR ratio $\mathcal{T}$ and the HE efficiency $\eta$. The figure indicates that the system can be operated in a parameter regime where the output power (current) is close to its maximal value, the efficiency approaches the Carnot bound $\eta_{\rm C} = 1$, and the TUR is significantly violated ($\mathcal{T} < 1$).

\begin{figure}
    \centering
  \includegraphics[width=1.0\columnwidth]{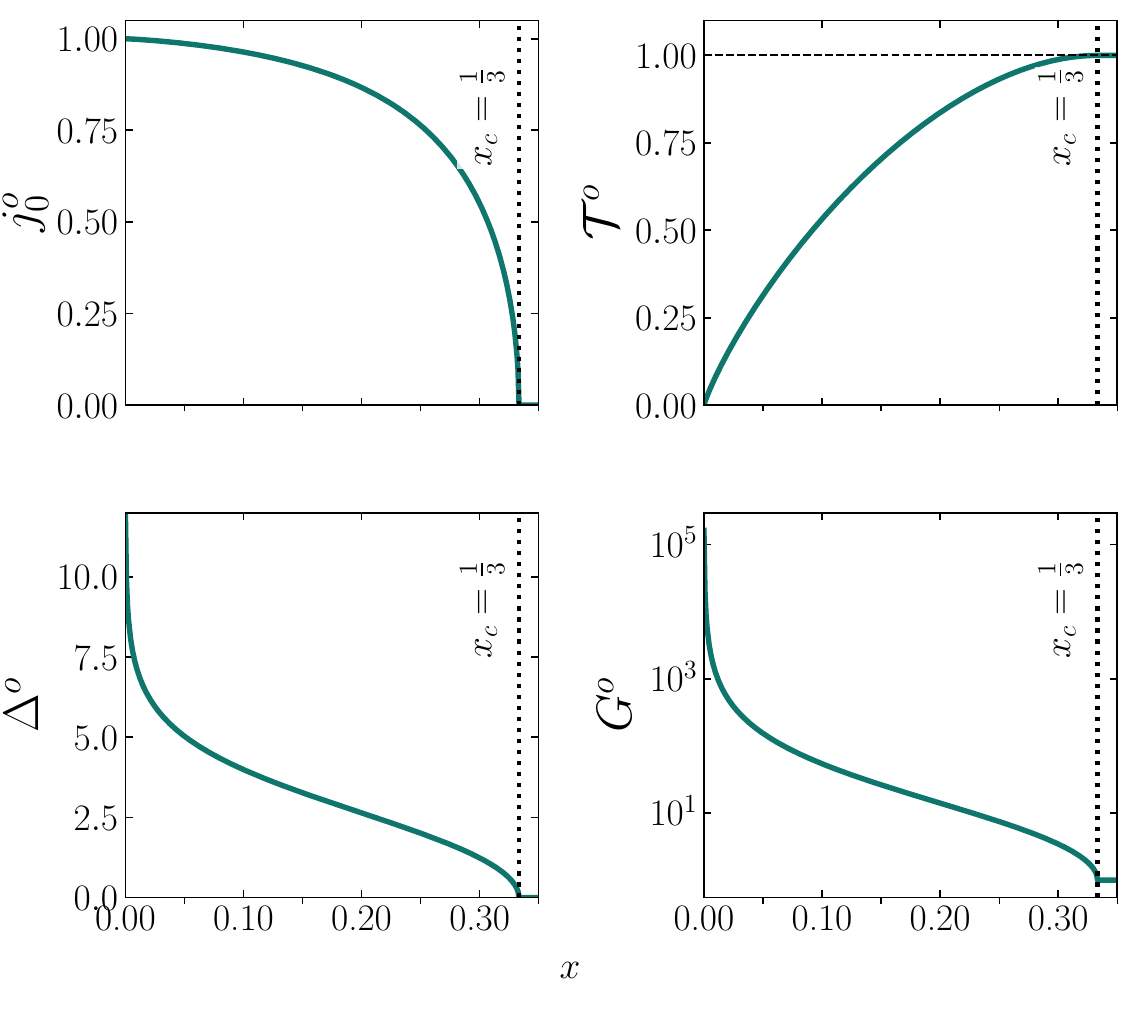}
  \caption{Optimal average displacement per half-oscillation, $j_0^{\mathrm{o}}$, TUR ratio, $\mathcal{T}^{\mathrm{o}}$, and the corresponding optimal cycle affinity, $\Delta^{\mathrm{o}}$, and degeneracy, $G^{\mathrm{o}}$, in the limit $T_h \to \infty$, shown as functions of the control precision (Fano factor) $x$. The value $x = 1/3$, at which the system operates reversibly and the TUR is no longer violated, is indicated by a vertical dotted line.   
}
   \label{fig:TURratchetTotalminima}
\end{figure}

In Fig.~\ref{fig:TURratchetminima}, we further show that the TUR ratio exhibits a minimum as a function of $\Delta$ for fixed $x > 0$. For $x = 0$, $\mathcal{T}^\star$ decreases monotonically to $0$, as discussed above.

Taking the derivative of $\mathcal{T}^\star$ in Eq.~\eqref{eq:Tstar_label} with respect to the affinity and setting it equal to zero, we obtain
\begin{equation}
\label{eq:Topt_stationary_IIIc}
\begin{split}
\frac{\partial \mathcal{T}^\star}{\partial \Delta}
&=\csch\Delta\bigl(1-\Delta\coth\Delta\bigr) \\
&\quad +x\left[\tanh\!\left(\frac{\Delta}{2}\right)
+\frac{\Delta}{2}\sech^2\!\left(\frac{\Delta}{2}\right)\right]=0 .
\end{split}
\end{equation}
We solve this transcendental equation numerically. The resulting optimal displacement $j_0^{\mathrm{o}}$, TUR ratio $\mathcal{T}^{\mathrm{o}}$, affinity $\Delta^{\mathrm{o}}$, and degeneracy $G^{\mathrm{o}}$ are shown as functions of the control accuracy in Fig.~\ref{fig:TURratchetTotalminima}.

For small $x$ (nearly deterministic control), the optimum shifts to large affinities: operating far from equilibrium suppresses backward transitions, thereby reducing relative fluctuations and enabling $\mathcal{T} < 1$. The TUR ratio vanishes for $x = 0$ and increases monotonically with increasing control noise until $x = x_c = 1/3$~\cite{Pietzonka2022}, where the system reaches the reversible limit with $\Delta = 0$, $j_0 = 0$, $\eta = \eta_{\rm C}$, and $\mathcal{T} = 1$.

Our analysis so far demonstrates that, in principle, the present HE can be operated simultaneously at maximum power (current) and vanishing TUR ratio when the cycle affinity and the hot-reservoir temperature diverge. Interestingly, in this parameter regime the HE efficiency can be tuned arbitrarily close to the Carnot efficiency, $\eta_{\rm C} = 1$. Indeed, choosing $E - F = \epsilon > 0$, the affinity in Eq.~\eqref{eq:deltaThinf} diverges for $T_c \ll \epsilon$, while the efficiency is given by
\begin{equation}
    \eta = \frac{1}{1 + \epsilon/(2F)}
    \approx 1 - \frac{\epsilon}{2F}.
\end{equation}
By selecting sufficiently small $T_c$ and $\epsilon$, while maintaining $T_c \ll \epsilon$, the HE efficiency can thus be made arbitrarily close to unity.

\section{Conclusions}
\label{sec:Conclusion}

We have introduced a minimal autonomous heat engine that strongly violates the long-time thermodynamic uncertainty relation. The model combines a discrete work-producing degree of freedom with an underdamped continuous mode that provides autonomous internal control. In the regime of time-scale separation, the dynamics becomes exactly solvable, allowing us to derive a closed analytical expression for the TUR ratio and to identify the mechanism responsible for its suppression. We have shown that the precision of the engine is governed by the Fano factor of the control-induced switching events: as the internal dynamics becomes increasingly deterministic, relative current fluctuations are strongly reduced while entropy production remains sufficiently small, enabling the TUR ratio to approach zero even near maximal current and efficiency.

Our results demonstrate that strong violations of thermodynamic uncertainty in autonomous HEs do not require complex potentials~\cite{CitalHolubec2026InertiaTames} or quantum architectures~\cite{Meier2025}, but can arise in a simple discrete setup coupled to an equilibrium underdamped harmonic oscillator.

Since the oscillator controlling the temperature and transition rates in the HE can be interpreted as an external drive, and since cyclic HEs are known not to obey the TUR and the PEC trade-off~\cite{HolubecMaxPower2018}, it may appear trivial that the present system also violates these bounds. However, this analogy breaks down once one recognizes that the notion of work in the two classes of systems differs qualitatively at the stochastic level. In the present system, work is performed through stochastic jumps against a bias, whereas in cyclic HEs it is generated by time modulation of the potential (in those systems, what corresponds here to current work is classified as heat). Moreover, the TUR is violated in entirely different parameter regimes: under strong nonequilibrium driving in the present setup, and in the reversible limit in cyclic HEs~\cite{HolubecMaxPower2018}.

In stationary underdamped models~\cite{Pietzonka2022,CitalHolubec2026InertiaTames}, including the present setup, the TUR is violated by exploiting the underdamped control degree of freedom to suppress current fluctuations, in the sense that both large and small power fluctuations are reduced. In the present model, this suppression is achieved by allowing transitions only between two discrete levels at a time, whereas in previously studied underdamped models it was realized through a carefully engineered potential landscape. 

In quantum systems, the TUR can be violated through nonclassical scaling of fluctuations when averaging over many correlated subsystems. In cyclic HEs, by contrast, the violation originates from the self-averaging property of work as defined in those systems. The mechanisms underlying TUR violation under these different dynamical conditions therefore appear to be fundamentally distinct.

We foresee two remaining challenges, one experimental and one theoretical. The experimental challenge is evident: to realize the predicted high-precision autonomous systems in practice. The theoretical challenge is more subtle: is there a universal principle underlying the violation of the TUR? In other words, can one construct a general framework that, based on these specific examples, yields a universal dynamical equation for the current in which an effective potential suppresses its fluctuations? If not, can such a framework at least be derived separately for each class of dynamical conditions?


\section*{Acknowledgements} EPC and VH were supported by Charles University (project PRIMUS/22/SCI/09 and GAUK grant 110-10/252750). We are grateful to Patrick Pietzonka for inspiring discussions, which initiated this project.

\bibliography{references} 

\end{document}